\RequirePackage[l2tabu,orthodox]{nag}
\documentclass[12pt]{article}
\pdfoutput=1
\usepackage{amsmath}
\usepackage{amssymb}
\usepackage{graphicx}
\usepackage{subfigure} 
\usepackage{url}
\usepackage{tikz}
\usepackage{authblk}
\usepackage[hyperfootnotes=false,hidelinks=true]{hyperref}
\title{More than one Author with different Affiliations}
\author[1,2]{Peter Wittek}
\author[3]{Shi Chao Gao}
\author[4]{Ik Soo Lim}
\author[3]{Li Zhao}
\affil[1]{University of Bor{\aa}s}
\affil[2]{ICFO-The Institute of Photonic Sciences}
\affil[3]{Tsinghua University}
\affil[4]{Bangor University}

\title{Somoclu: An Efficient Parallel Library for Self-Organizing Maps}
\date{}
\begin{document}
\maketitle

\begin{abstract}Somoclu is a massively parallel tool for training
  self-organizing maps on large data sets written in
  C++. It builds on OpenMP for multicore execution,
  and on MPI for distributing the workload across the nodes
  in a cluster. It is also able to boost training by using CUDA
  if graphics processing units are available. A sparse kernel is
  included, which is useful for high-dimensional but sparse data, such
  as the vector spaces common in text mining
  workflows. Python, R and MATLAB
  interfaces facilitate interactive use. Apart from fast execution,
  memory use is highly optimized, enabling training large emergent
  maps even on a single computer.
\end{abstract}

\section{Introduction}
Visual inspection of data is crucial to gain an intuition of the
underlying structures. As data often lies in a high-dimensional space,
we use embedding techniques to reduce the number of dimensions to just
two or three.

Methods that rely on eigenvalue decomposition, such as
multidimensional scaling~\cite{cox1994mds}, achieve a global optimum
for such an embedding: The global topology of the space will be
preserved.

Often the data points lie on a high-dimensional manifold that is
curved and nonlinear. These structures are difficult to find with
eigenvalue decomposition. Manifold learning generalizes embedding
further, assuming that data in a high-dimensional space aligns to a
manifold in a much lower dimensional space. For example, the algorithm
Isomap finds a globally optimal solution for an underlying nonlinear
manifold and it builds on multidimensional
scaling~\cite{tenenbaum2000global}. Isomap, however, fails to find
nonconvex embeddings~\cite{weinberger2004learning}.

Nonconvex structures are one strong motivation to look at solutions
that are not globally optimal, but preserve the local topology of the
space instead. Self-organizing maps (SOMs) are a widespread
visualization tool that embed high-dimensional data on a
two-dimensional surface -- typically a section of a plane or a torus
-- while preserving the local topological layout of the original
data~\cite{kohonen2001som}. These maps provide a visual
representation of groups of similar data instances, and they are also
useful in analyzing the dynamics of evolving data collections, as
updates are easily made. Emergent self-organizing maps contain a much
larger number of target nodes for embedding, and thus capture the
topology of the original space more accurately~\cite{ultsch2005esom}.

Training a map is computationally demanding, but a great advantage of
SOMs is that the computations are easy to parallelize. They have been
accelerated on massively parallel graphics processing units
(GPUs~\cite{luo2005self}). Tools exist that scale to large data sets
using cluster resources~\cite{sul11sommr}, and also combining
GPU-accelerated nodes in clusters~\cite{wittek2012gpusom}. These
tools focus on batch processing. On the other end of the spectrum,
interactive environments for training SOMs are often single-core
implementations~\cite{ultsch2005esom}, not making full use of
contemporary hardware.

We develop a common computational core that is highly efficient with
the available resources. This common core is used for a command-line
interface to enable batch processing on cluster resources, but the
same core is used as the computational back-end for environments
popular in data analysis. The tool, named Somoclu (originally
standing for SOM on a cluster), has the following improvements over
other implementations:

\begin{itemize}
\item It is highly efficient in single-node multicore execution.
\item Memory use is reduced by up to 50 per cent.
\item Large emergent maps are feasible.
\item A kernel for sparse data is introduced to facilitate text mining applications.
\item Training time is reduced by graphics processing units when available.
\item It improves the efficiency of distributing the workload across multiple nodes when run on a cluster.
\item An extensive command-line interface is available for batch processing.
\item Python~\cite{vanrossum2014python},
  R~\cite{rcoreteam2013rinstallation} and
  MATLAB~\cite{mathworks2011matlab} interfaces facilitate
  interactive processing.
\item Compatibility with Databionic ESOM Tools~\cite{ultsch2005esom} ensures easy visualization.
\end{itemize}
The source code is available under GNU Public License Version 3
(\url{http://peterwittek.github.io/Somoclu/}). The Python
version is also listed in the Python package index
(\url{https://pypi.python.org/pypi/Somoclu/}). The R
version is available from the Comprehensive R Archive Network
(CRAN) at \url{https://CRAN.R-project.org/package=RSomoclu/}.

\section{Self-organizing maps in batch mode}
The SOM training algorithm constructs a nonlinear topology preserving
mapping of the input data set
$X = \{x(t)|t\in \{t_{0, \ldots, t_{f}}\}\}$, where $t_{0}$ and
$t_{f}$ are the beginning and the end of the current training session,
onto a set of neurons $M=\{n_1, \ldots, n_k\}$ of a neural
network~\cite{kohonen2001som}. The neurons of the network are
arranged in a grid, with associated weight vectors
\begin{equation}
W = \{w_1(t), \ldots, w_k(t)\}
\end{equation}
at a given time step $t$. $W$ is known as the code book.

Each data point $x(t)$ is mapped to its best matching unit
\begin{equation}
  \mathrm{bm}(x(t))= n_b\in  M,
\label{bmu}
\end{equation}
such that
\begin{equation}
d(x(t), w_b(t))\leq d(x(t), w_j(t)) \quad \forall w_j(t)\in W,
\end{equation}
where $d$ is the distance function on the data set in the feature
space. The neurons are arranged on a two dimensional map: Each neuron
$i$ has a two coordinates in a grid. Next the weight vector of the
best match neuron and its neighbors are adjusted toward the input
pattern using the following equation:
\begin{equation}
w_j(t+1)=w_{j}(t)+\alpha h_{bj}(t)(x(t)-w_{j}(t)),
\end{equation}
where $0<\alpha<1$ is the learning factor, and $h_{bj}(t)$ is the neighborhood function that decreases for neurons further away from the best match neuron in grid coordinates. A frequently used neighborhood function is the Gaussian:
\begin{equation}
  h_{bj}=\exp\left(\frac{-\|r_{b}-r_{j}\|}{\delta(t)}\right),
\end{equation}
where $r_{b}$ and $r_{j}$ stand for the coordinates of the respective nodes. The width $\delta(t)$ decreases from iteration to iteration to narrow the area of influence.

The training is repeated on the same data set to increase the fit, a training cycle is called an epoch. Eventually, the neighborhood function decreases to an extent that training might stop. The time needed to train an SOM grows linearly with the data set size, and it grows linearly with the number of neurons in the SOM. SOM has a batch formulation of updating the weights, which is widely used in parallel implementations:
\begin{equation}
w_j(t_{f})=\frac{\sum_{t'=t_{0}}^{t_{f}}h_{bj}(t')x(t')}{\sum_{t'=t_{0}}^{t_{f}}h_{bj}(t')}.
\label{batch}
\end{equation}

While not directly related to the training, it is worth mentioning the U-matrix associated with a SOM, which depicts the average Euclidean distance between the code book vectors of neighboring neurons. Let $N(j)$ denote the immediate neighbors of a node $j$. Then the height value of the U-matrix for a node $j$ is calculated as
\begin{equation}
  U(j)=\frac{1}{|N(j)|}\sum_{i\in N(j)} d(w_i,w_j).
\end{equation}
The purpose of the U-matrix is to give a visual representation of the topology of the network.

There is no shortage of implementations of SOM training algorithms. It
comes integrated in data mining suites such as
RapidMiner~\cite{hofmann2013rapidminer}. Dedicated
visualization and training tools also exist, such as the
Databionic ESOM Tools~\cite{ultsch2005esom}. Popular languages
used in data analytics all have SOM modules, including
MATLAB~\cite{vesanto1999self},
Python~\cite{hanke2009pymvpa}, and
R~\cite{wehrens2007self}. Common to these tools is that
they seldom make use of parallel computing capabilities, although the
batch formulation of SOM training invites such
implementations. Further constraints include memory requirements,
which increase fast with the size of the neuron grid.

Lower level libraries address such scalability issues. Distributing
the workload across multiple nodes is feasible via standard methods
such as MPI and
MapReduce~\cite{lawrence1999scalable,sul11sommr}. The inherent
parallelism is easy to exploit with graphics processing units
(GPUs~\cite{luo2005self}). A method combining distributed
computing with parallelism on GPUs is also
available~\cite{wittek2012gpusom}.

With Somoclu, we bridge high-level languages and massively
parallel architectures, while also simplifying the distribution of
workload compared to earlier implementations. We use a common core
written in C++, supplemented by
OpenMP~\cite{dagum1998openmp}, MPI~\cite{snir1998mpi},
and CUDA~\cite{nvidia2014cuda}. This common core is used by an
extensive command-line interface, and wrappers in Python,
R, and MATLAB.

\section{High-performance core for training}
The starting point was a MapReduce-based distributed implementation of SOM~\cite{sul11sommr}. This implementation was later accelerated by GPUs~\cite{wittek2012gpusom}. We found the MapReduce framework unnecessary, and derived a purely MPI-based implementation, that is more modular, faster, and includes a sparse kernel. Design details are provided in Section~\ref{design}. Distributing the workload across multiple nodes is an extension of the parallel formulation (Section~\ref{distributed}).

\subsection{Parallelism}\label{design}
\begin{figure}[t!]
\centering
\begin{tikzpicture}
\draw [rounded corners] (-0.2,2.8) rectangle (8.5,10.2);
\node [above right] at (-0.2,2.8) {Process};

\draw [fill=lightgray,rounded corners] (0,3.5) rectangle (4.4,10);
\node [above right] at (0,3.5) {Memory structures};

\path [fill=gray] (0.2,6.7) rectangle (4.2,9.8);
\node [above right] at (0.2,6.7) {Code book};

\path [fill=gray] (0.2,5) rectangle (4.2,6.5);
\node [above right] at (0.2,5) {Data chunk};

\path [fill=gray] (0.2,4.1) rectangle (4.2,4.8);
\node [above right] at (0.2,4.1) {Best matching units};

\draw [fill=lightgray,rounded corners] (5,3.5) rectangle (8.2,10);
\node [above right] at (5,3.8) {OpenMP/CUDA};
\node [above right] at (5,3.5) {threads};

\foreach \y in {4.5,5,...,9.8}
{
  \draw [fill=darkgray] (5.2,\y) rectangle (8,\y+0.3);
}
\end{tikzpicture}
\caption{Overview of the parallel organization of the batch training. The global code book is replicated in each process, therefore it is more efficient to use OpenMP-based parallelization than to rely on MPI on multicore CPUs. If a GPU is available, it will overtake most data parallel operations from the CPU, replacing OpenMP threads.}
\label{parallelorganization}
\end{figure}
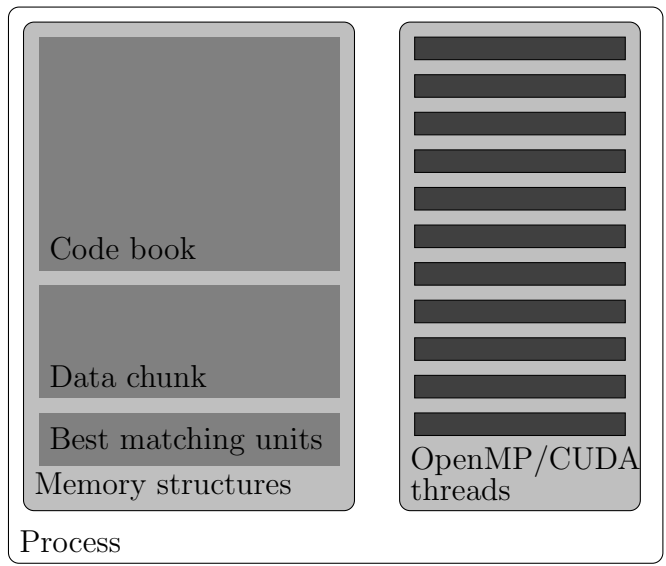

The dense CPU kernel is a straightforward implementation of the batch
formulation in Equation~\ref{batch}, and it resembles the one
implemented by Ref.~\cite{sul11sommr}. For an overview of the parallel
organization, refer to Figure~\ref{parallelorganization}. The
MapReduce calls were replaced by MPI functions, leading to a
more streamlined code. Furthermore, single-node parallelism on
multicore CPUs is redesigned to use OpenMP instead of
MPI. This strategy avoids duplicating the code book: Each
MPI process must have a copy of the code book, but OpenMP
threads can work on the same copy. The simplification leads to a
minimum fifty per cent reduction in memory even when only two threads
are used.

A performance bottleneck in the original implementation was the accumulation of local weights into a new global code book by one single process on the master node. This is parallelized by an OpenMP directive. Furthermore, the influence radius $\delta(t)$ of a best matching node is thresholded, which translates to speed improvements without compromising the quality of the trained map.

The GPU variant is more complex compared to the CPU kernel. The complexity stems from the way the distance function is evaluated between the nodes of the SOM and the training data. To maximize parallelism, the Gram matrix is calculated, that is, a matrix of the distances between every data instance and the nodes of the SOM. A na\"ive approach would be to extend an efficient matrix multiplication algorithm, replacing the dot product by the distance function. Opting for a Euclidean distance, it is possible to derive an alternative formulation of calculating the Gram matrix using linear algebra operations~\cite{li2011chunking}. Benchmarking the two approaches, we found that the latter approach is a magnitude faster on the GPU, mainly due to a more favorable memory access pattern.

We implemented the GPU kernel with
Thrust~\cite{bell2011thrust}, a C++ template library
for CUDA, which has high-performance primitives, avoiding the
need to manually tune individual GPU calls. Compared to the
implementation by Ref.~\cite{wittek2012gpusom}, the device memory use of
the GPU code is reduced to approximately one-third, and the new
implementation completely avoids costly matrix transposing operations.

Further complexity arises from the disparity between the number of
GPUs and the number of cores in a computer. The CPU kernel achieves
maximum speed by running an MPI process on each available core,
resulting in a far lower number of data instances per core and a
speedier local update of the weights. For instance, if there are eight
cores and two GPUs, than each GPU has four times more data to process
and its corresponding MPI thread would have four times more data
to update the local weights. While the GPU handles the load
efficiently, it would be highly inefficient to use a single thread to
update the local weights. We thus hybridized the kernel and rely on
OpenMP to parallelize the weight update. The GPU implementation
runs as many MPI processes on a node as there are GPUs, and uses
all CPU cores in the weight update.

The sparse kernel is a straightforward extension of the dense CPU
kernel, and its main virtue is the reduced memory use. A vector space
coming from a text processing pipeline typically contains 1--5\%
nonzero elements, leading to a 20--100$\times$ reduction in memory use
when using a sparse representation. This kernel does not have a GPU
implementation, as the irregular access patterns that are inevitable
with sparse data structures are not efficient on streaming
architectures.

\subsection{Workload in distributed environment}\label{distributed}
The training procedure has a simple communication structure in a distributed memory system. Finding the best matching unit in Equation~\ref{bmu} is independent for every data instance, resembling the parallel version. This means that we can distribute equally sized parts of the data to each node, without any further communication of training data later on.

The update of the code book requires two way communication between a
master node and the slave nodes. Once all weight updates are
calculated in a slave node, the local updates are sent to the master
node, which accumulates the changes to the code book. The new code
book is broadcast to all slave nodes.

This communication structure ensures a near-linear scaling for smaller code books. Emergent maps with high-dimensional data, however, scale poorly. The code book is always a dense structure, even if the training data is sparse, as there are hardly any zero entries. Storing the code book in memory is the primary constraint for single node execution, and it is also the key constraint in distributed workloads.

Compared to an earlier GPU-accelerated, distributed implementation of SOM~\cite{wittek2012gpusom} that used MapReduce, we found that ordinary MPI calls are sufficient. Execution time did not improve significantly, but the overall code simplified, and our implementation does not have to rely on a little-used, MPI-based MapReduce library.

\section{Interfaces}
We regard the package libsvm for training support vector
machines as a role model: Using an efficient computational core,
numerous wrappers were added to the library to interface with other
languages and environments~\cite{chang01}. The package
fastcluster is also similar in its
approach~\cite{mullner2013fastcluster}.

We provide an extensive command-line interface, which is able to
access all functionality provided by the computational core
(Section~\ref{usage}). The training functionality is exposed via an
application programming interface, a single function that performs one
epoch of training (Section~\ref{api}). This function is wrapped to be
called from Python, R, and MATLAB
without duplicating data structures (Section~\ref{interfaces}).

\subsection{Command-line interface}\label{usage}
One sparse and two dense data formats are supported. All of them are plain text files. The entries can be separated by any white-space character. One row represents one data instance across all formats. Comment lines starting with a hash mark are ignored.

The basic dense format includes the coordinates of the data vectors, separated by a white-space. This file is parsed twice to get the basic dimensions right. The second dense format is identical, but it includes a header that contains information about the layout of the matrix. This second format is compatible with Databionic ESOM Tools.

The sparse representation is similarly row-oriented, and it uses the
same sparse format as libsvm~\cite{chang01}. For instance, the
vector [1.2 0 0 3.4] is represented as the following line in the
file: 0:1.2 3:3.4. The file is parsed twice: once to get the number of
instances and features, and the second time to read the data in the
individual threads.

Example files are available in the package. Extension to further matrix types is simple to add.

The tool is used via the command line. Somoclu takes a plain
text input file -- either dense or sparse data. The basic execution is
as follows:

\begin{verbatim}
$ [mpirun -np NPROC] Somoclu [OPTIONs] INPUT_FILE OUTPUT_PREFIX
\end{verbatim}

The \verb+mpirun+ is required only when multiple nodes are used, or when there is more than one GPU in the system. The parameter \verb+INPUT_FILE+ is self-explanatory. Instead of names of output files for the best matching units, code books, and U-matrices, an output prefix is requested in \verb+OUTPUT_PREFIX+. The resulting files will be differentiated by the extension, and, if interim snapshots are requested, also by the indices of the epochs in which the snapshots are taken.

The rest of the arguments are as follows.

\begin{verbatim}
-c FILENAME
\end{verbatim}

The parameter specifies an initial code book for the map. The default is random initialization.

\begin{verbatim}
-e NUMBER
\end{verbatim}

The number of training epochs, that is, the number of times each data instances will be presented to the learner.

\begin{verbatim}
-g TYPE
\end{verbatim}

The type of the grid type; the default is square and the other option is hexagonal.

\begin{verbatim}
-k NUMBER
\end{verbatim}

This parameter defines the kernel type. The value 0 stands for the dense CPU kernel, 1 is for the dense GPU kernel, and 2 is for the sparse CPU kernel.

\begin{verbatim}
-m TYPE
\end{verbatim}

The map type is either planar or toroid; the default is planar.

\begin{verbatim}
-n FUNCTION
\end{verbatim}

This option chooses the neighborhood function, which can be Gaussian or bubble; the default being the former.

\begin{verbatim}
-p NUMBER
\end{verbatim}

If the neighborhood function is Gaussian, setting this parameter to 1 will cut off the update of nodes beyond the current radius.

\begin{verbatim}
-t STRATEGY
\end{verbatim}

Radius cooling strategy is either linear or exponential; the default is linear.

\begin{verbatim}
-r NUMBER
\end{verbatim}

The parameter defines the start radius. The default value is half of the map size in the smaller direction.

\begin{verbatim}
-R NUMBER
\end{verbatim}

The final radius; it defaults to 1.

\begin{verbatim}
-T STRATEGY
\end{verbatim}

Learning rate cooling strategy is either linear or exponential; the default is linear.

\begin{verbatim}
-l NUMBER
\end{verbatim}

The starting learning rate; with a default value of 1.0.

\begin{verbatim}
-L NUMBER
\end{verbatim}

The final learning rate; the default is 0.01.

\begin{verbatim}
-s NUMBER
\end{verbatim}

This parameter decides whether to save interim files. The default
value is 0, which means that no interim files will be saved. Setting
the parameter to 1 will calculate and save the U-matrix in each time
step. If the value passed is 2, then the code book and the best
matching units are also saved in each epoch. Since the code book is
enormous for large emergent maps, passing a value of 2 may
significantly slow down each epoch.

\begin{verbatim}
-x, --columns NUMBER
\end{verbatim}

This is the number of columns in map, that is, the size of the SOM in
direction $x$; the default value is 50.

\begin{verbatim}
-y, --rows    NUMBER
\end{verbatim}

The number of rows in map is defined by this value, that is, the size
of the SOM in direction $y$; the default value is 50.  Examples:

\begin{verbatim}
$ Somoclu data/rgbs.txt data/rgbs
$ mpirun -np 4 --hostfile hostfilename Somoclu -k 0 --rows 20 --columns 20 \
>   data/rgbs.txt data/rgbs
\end{verbatim}

%
\subsection{As an application programming interface}\label{api}
Designed to work with MPI, Somoclu was not conceived to be
used as an API. Yet, given sufficient preparation on the calling side,
it is possible to interface with Somoclu as an API. The primary
entry point for training is the following function, as specified in
\verb+Somoclu.h+:

\begin{verbatim}
void trainOneEpoch(int itask, float *data, svm_node **sparseData,
  float *codebook, int *globalBmus, unsigned int nEpoch,
  unsigned int currentEpoch, unsigned int nSomX, unsigned int nSomY,
  unsigned int nDimensions, unsigned int nVectors,
  unsigned int nVectorsPerRank, unsigned int radius0,
  unsigned int radiusN, string radiusCooling, float scale0, float scaleN,
  string scaleCooling, unsigned int kernelType, string mapType,
  string gridType, bool compact_support, bool gaussian)
\end{verbatim}

The parameters \verb+nSomX+, \verb+nSomY+, \verb+nEpoch+, \verb+kernelType+, and so on, are self-explanatory, they are the same as in the command-line interface.

The parameter \verb+itask+ specifies the rank of the current MPI process. If the calling environment does not use MPI, the value should be set as zero.

Data are either stored in \verb+data+ or \verb+sparseData+, depending on whether the data are dense or sparse. Only data belonging to the process is expected, not the full matrix. The parameter \verb+nVectors+ stores the total number of instances, whereas \verb+nVectorsPerRank+ the number of instances belonging to one MPI thread.

\subsection[Python, R and MATLAB interfaces]{Python, R and MATLAB interfaces}\label{interfaces}
Solutions for scalable data processing in higher level languages tend
to be ad hoc. Focusing on R as an example, the R
Installation and Administration manual states that the language is
ill-suited for working with data structures larger than about 10--20\%
of the main memory~\cite{rinstallation}. The key problem is with the
overheads on the data structures. Parallel processing is scarcely
addressed. Some modules aim to overcome these
barriers~\cite{kane2013scalable}.

Distributed computing is also challenging for such languages. Most
solutions work entirely on a case by case basis, such as the
distributed text mining module for
R~\cite{theussl2012tmplugin}.

The aim of the interfaces is to address the problem of memory use and
parallel execution. The Python package easily installable
from the Python package index. The R package is available
from CRAN. These packaged variants rely on the OpenMP parallelism
alone. To get the massive CUDA parallelism through the
interfaces, users with basic software building knowledge can easily
follow the instructions on the project site. MATLAB users
need to follow the instructions on the project site to build the
module as no central package distribution site is
available. Distributed workloads directed from a high-level language
remain for future development.

Starting with the Python interface, our primary concern is
to avoid memory copies and duplicating large structures in memory. The
native variable system of Python is not strictly typed,
hence we require the use of numpy. Using numpy
\verb+float32+ data type for storing arrays, we map directly to the
single-precision floats used in the C++ code. We pass
pointers between the two languages, making the best use of memory and
almost zero computational overhead. This interface encapsulates the
training routines in a class to simplify the workflow.

An example use of the interface is as follows:

\begin{verbatim}
>>> import Somoclu
>>> import numpy
>>> data = numpy.loadtxt('data/random.dat')
>>> n_rows, n_columns = 50, 50
>>> som = Somoclu.Somoclu(n_columns, n_rows, data = data)
>>> som.train()
\end{verbatim}

The final code book is in the class variable \verb+Somoclu.codebook+,
the best matching units are in \verb+Somoclu.bmus+, and the U-matrix
is also calculated in \verb+Somoclu.umatrix+.

The R interface uses the Rcpp package to cast
pointers and provide type conversions between R and
C++~\cite{eddelbuettel2011rcpp}. Since R uses
double precision matrices by default, and Somoclu uses
single-precision floats internally, we must convert between double and
float arrays of the input and output. This makes the interface less
efficient than the Python version. The interface integrates
with the popular R SOM package kohonen to facilitate
use. The following is an example on how to call the wrapper function
from R:

\begin{verbatim}
R> library("RSomoclu")
R> data("rgbs", package = "RSomoclu")
R> input_data <- data.matrix(rgbs)
R> nSomX <- 50; nSomY <- 50
R> nEpoch <- 10
R> radius0 <- 0; radiusN <- 0
R> radiusCooling <- "linear"
R> scale0 <- 0; scaleN <- 0.01
R> scaleCooling <- "linear"
R> kernelType <- 0
R> mapType <- "planar"
R> res <- RSomoclu.train(input_data, nEpoch, nSomX, nSomY, radius0, radiusN,
+    radiusCooling, scale0, scaleN, scaleCooling, kernelType, mapType)
\end{verbatim}

The code book is returned in
\verb+res$codebook+, the best matching units are in
\verb+res$globalBmus+, and the calculated final U-matrix is in
\verb+res$uMatrix+. Most R users will probably want to work
with the map through the package kohonen:

\begin{verbatim}
R> sommap <- RSomoclu.kohonen(input_data, res)
\end{verbatim}

The MATLAB interface uses the official MEX-file mechanism
to interface with C++. As with R,
MATLAB also defaults to double precision variables,
rendering some overhead of type conversion inevitable. We designed a
similar training function call to that of the SOM Toolbox
\cite{vesanto1999self}. And provided examples on visualization using
both SOM Toolbox and Databionic ESOM Tools. The following
is an example call from the MATLAB interface.

\begin{verbatim}
>> data = importdata('data/random.dat');
>> msize = [50 50];
>> sMap  = som_randinit(D, 'msize', msize);
>> nEpoch = 10;
>> radius0 = 0;
>> radiusN = 0;
>> radiusCooling = 'linear';
>> scale0 = 0;
>> scaleN = 0.01;
>> scaleCooling = 'linear';
>> kernelType = 0;
>> mapType = 'planar';
>> gridType = 'rectangular';
>> compactSupport = false;
>> neighborhood = 'gaussian';
>> [sMap, sTrain, globalBmus, uMatrix] = ...
  Somoclu_train(sMap, data, 'msize', msize, 'radius0', radius0, ...
  'radiusN', radiusN, 'radiusCooling', radiusCooling, ...
  'scale0', scale0, 'scaleN', scaleN, 'scaleCooling', scaleCooling, ...
  'kernelType', kernelType, 'mapType', mapType, ...
  'gridType', gridType, 'compactSupport', compactSupport, ...
  'neighborhood', neighborhood, 'nEpoch', nEpoch);
\end{verbatim}

The returned variables are identical to the Python version.

\subsection{Visualization}\label{visualization}
The primary purpose of generating a map is visualization. Yet -- apart
from the Python interface -- Somoclu does not come
with its own functions for visualization, since there are countless
generic and SOM-specific tools that are capable of plotting
high-quality figures. The code book, the best matching units, and the
U-matrix are exported at the end of the training, or even after each
epoch.

The simplest procedure is to use a generic plotting library, such as gnuplot. For instance, the following simple gnuplot script plots the U-matrix in PNG format:

\begin{verbatim}
gnuplot> set datafile commentschars "#!%"
gnuplot> set autoscale
gnuplot> unset log
gnuplot> unset label
gnuplot> set lmargin at screen 0
gnuplot> set rmargin at screen 1
gnuplot> set bmargin at screen 0
gnuplot> set tmargin at screen 1
gnuplot> unset xtic
gnuplot> unset ytic
gnuplot> set notitle
gnuplot> set pm3d at b
gnuplot> set pm3d map
gnuplot> set nokey
gnuplot> set nocolorbox
gnuplot> unset surface
gnuplot> set term pngcairo transparent size 500,500
gnuplot> set output 'umatrix.png'
gnuplot> splot "filename"  matrix
\end{verbatim}

The dimensions of the picture should be adapted to the size of the network to avoid distortion.
\begin{figure}[t!]
  \centering
  \subfigure[The complete map.]{
  \includegraphics[width=0.47\textwidth, trim = 0 8 0 42, clip]{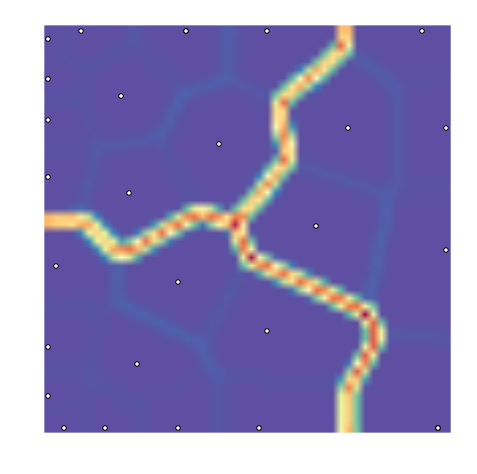}
  }
  \subfigure[A close-up of region.]{
  \includegraphics[width=0.47\textwidth, trim = 0 8 0 42, clip]{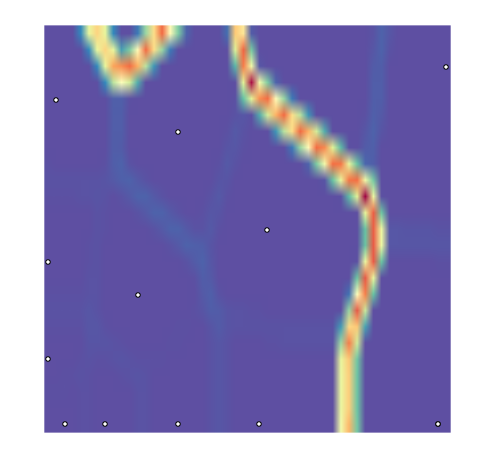}
  }
  \caption{The U-matrix of a toroid map with labels trained on a toy example plotted in Python.}
  \label{figure0}
\end{figure}

The Python interface is equipped with plotting
routines. For instance, a map can be plotted along with the labels or
class colors:

\begin{verbatim}
>>> som.view_umatrix(bestmatches = True)
\end{verbatim}

We may also zoom in to a specific part of the map:

\begin{verbatim}
>>> som.view_umatrix(bestmatches = True, zoom = ((n_rows//2, n_rows), (0, n_columns)))
\end{verbatim}

Using a toy example included with the code, we obtained a visualization shown in Figure~\ref{figure0}.

\begin{figure}[t!]
  \centering
  \subfigure[Codebook.]{
  \includegraphics[width=0.47\textwidth, trim = 0 8 0 36, clip]{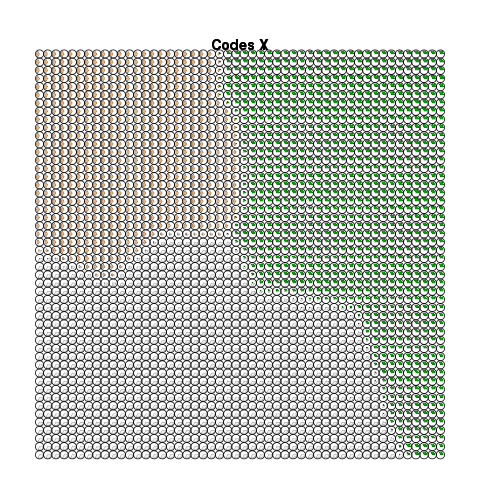}
  }
  \subfigure[U-matrix.]{
  \includegraphics[width=0.47\textwidth, trim = 0 8 0 36, clip]{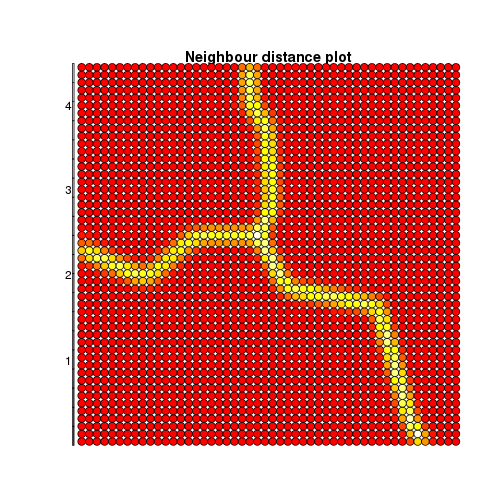}
  }
  \caption{Visualizing a toy example trained by Somoclu and
    plotted by kohonen in R.}
  \label{figure0R}
\end{figure}

The R interface allows visualization through the package
kohonen. This is entirely external to the library. A conversion
to kohonen's format is necessary:

\begin{verbatim}
R> sommap <- RSomoclu.kohonen(input_data, res)
R> plot(sommap, type = "codes", main = "Codes"))
R> plot(sommap, type = "dist.neighbours")
\end{verbatim}

A toy example is shown in Figure~\ref{figure0R}.

The results of the MATLAB interface allows visualization through the SOM Toolbox, which is also external to the library.

\begin{verbatim}
>> som_show(sMap);
\end{verbatim}
\begin{figure}[t!]
  \centering
  \includegraphics[width=\textwidth, trim = 0 5 0 20, clip]{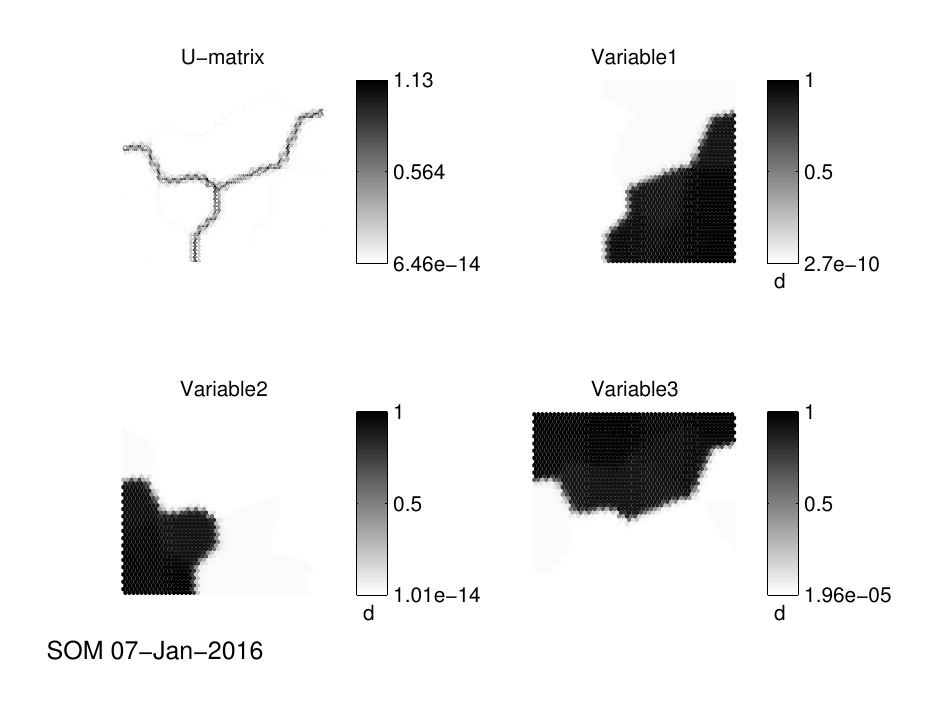}
  \caption{Visualizing a toy example trained by Somoclu and plotted by SOM Toolbox in MATLAB.}
  \label{figureMATLAB}
\end{figure}

The toy example is shown Figure~\ref{figureMATLAB}.

More advanced, SOM-specific tools may also offer an interactive GUI to
select gradients and other parameters. One such tool is
Databionic ESOM Tools~\cite{ultsch2005esom}, with which the
output formats of Somoclu are compatible. The result of such
visualization is shown in Section~\ref{visualizationresult}.
\section{Experimental results}
To ensure replicability of the results, we benchmarked with publicly available cluster GPU instances provided by Amazon Web Services. The instance type was cg1.4xlarge,\footnote{\url{https://aws.amazon.com/ec2/instance-types/}} equipped with 22 GiB of memory, two Intel Xeon X5570 quad-core CPUs, and two NVIDIA Tesla M2050 GPUs, running Ubuntu 12.04.

\begin{figure}[t!]
\centering
\subfigure[$50{\times}50$ map.]{
\includegraphics[width=0.47\textwidth, trim = 0 5 0 10, clip]{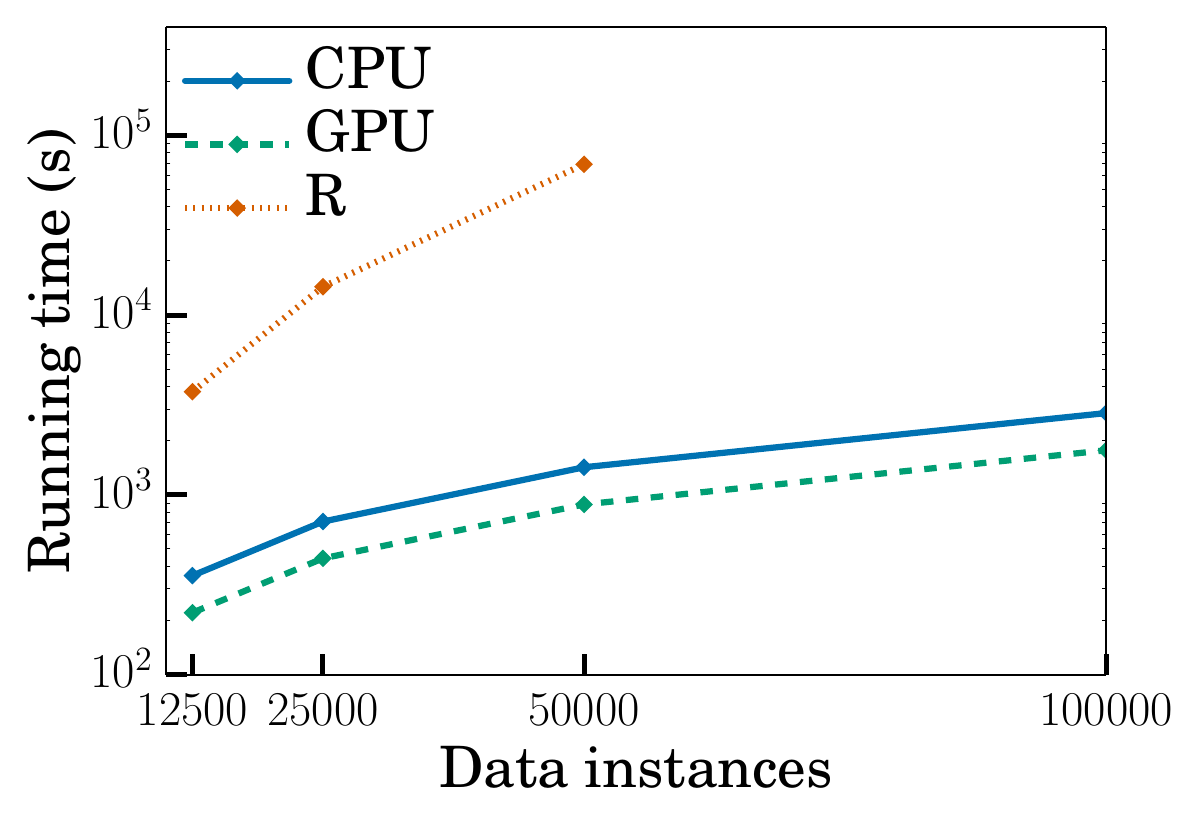}
\label{fig:subfig1}
}
\subfigure[$200{\times}200$ emergent map.]{
\includegraphics[width=0.47\textwidth, trim = 0 5 0 10, clip]{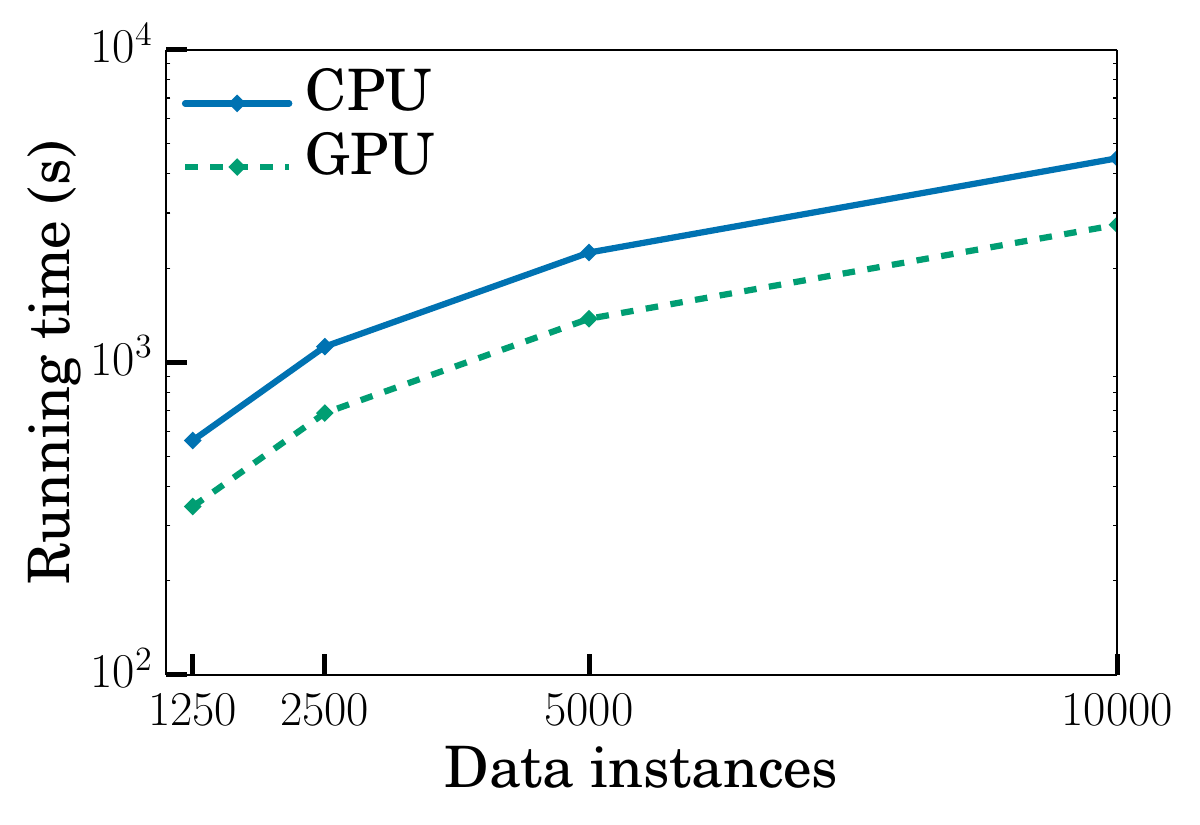}
\label{fig:subfig2}
}
\caption{Training time on a single node with CPU and GPU kernels and the R package kohonen. The time axis is logarithmic. The data instances had 1,000 dimensions.}
\label{figure1}
\end{figure}
\subsection{Single-node performance}

The most direct comparison should be with the implementations
by Refs.~\cite{sul11sommr,wittek2012gpusom}. Unfortunately the
former was not able to handle the matrix sizes we were benchmarking
with. The implementation by Ref.~\cite{wittek2012gpusom} is sensitive to
the size of map, and it did not scale to the map sizes benchmarked
here, and thus we left it out from the comparison. To compare with
single-core performance, we included the R package
kohonen~\cite{wehrens2007self}. The number of data instances
ranged from 12,500 to 100,000, the number of dimensions was fixed at
1,000 for a regular $50\times 50$ self-organizing map. The data
elements were randomly generated, as we were interested in scalability
alone. We also tested an emergent map of $200\times{}200$ nodes, with
the number of training instances ranging from 1,250 to 10,000. This
large map size with the largest data matrix filled the memory of a
single GPU, hence giving an upper limit to single-node
experiments. Emergent maps in the package kohonen are not
possible, as the map is initialized with a sample from the data
instances. If the map has more nodes than data instances,
kohonen exits with an error message. The map size did not affect
the relative speed of the different kernels
(Figure~\ref{figure1}). The comparison was based on the command-line
version of Somoclu.

\begin{figure}[t!]
\centering
\subfigure[Running time.]{
  \includegraphics[width=0.47\textwidth, trim = 0 5 0 10, clip]{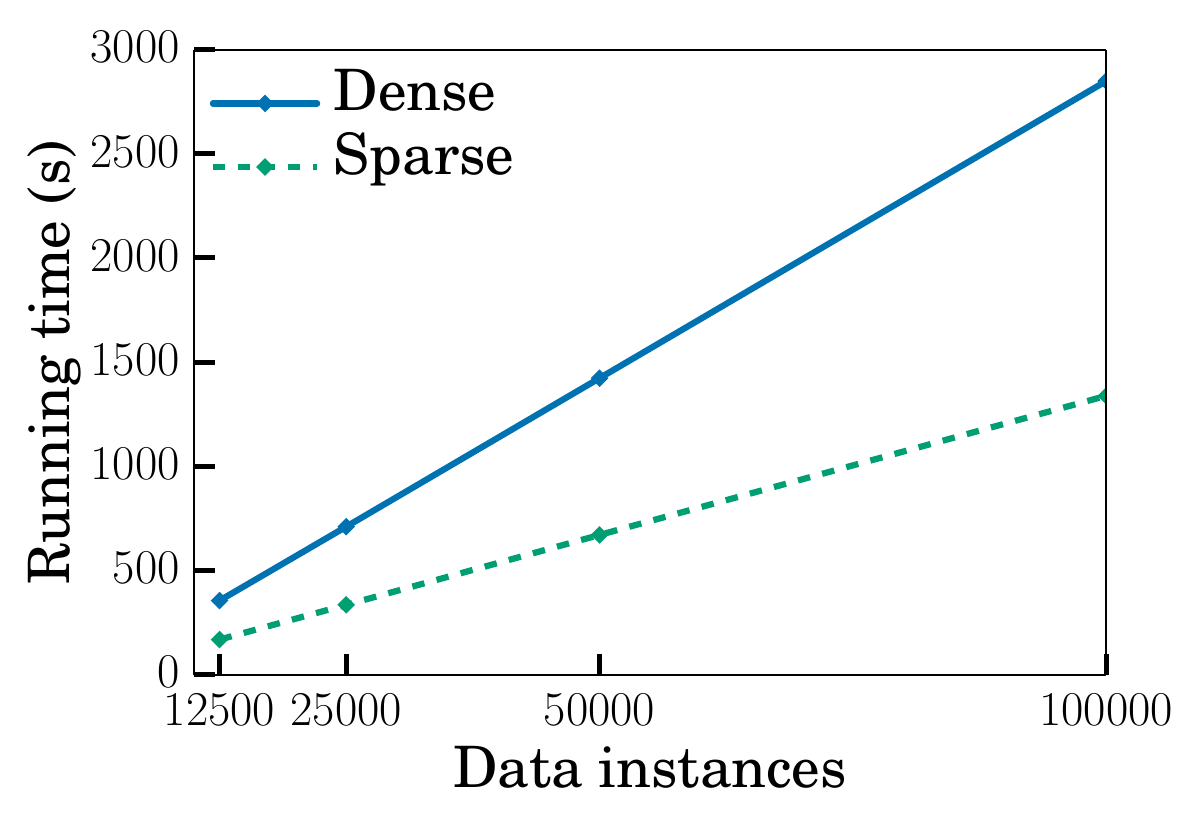}
\label{sparserunningtime}
}
\subfigure[Memory use.]{
\includegraphics[width=0.47\textwidth, trim = 0 5 0 10, clip]{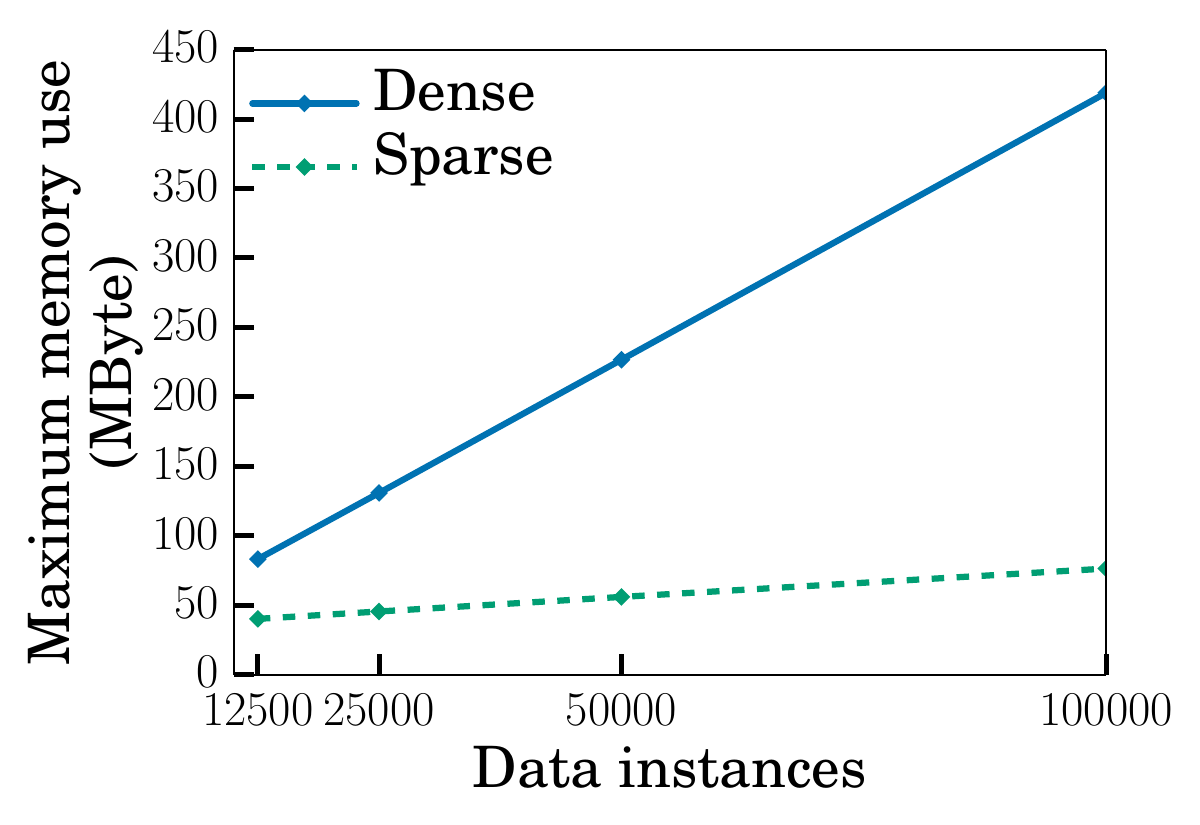}
\label{sparsememory}
}
\caption{Training time on a single node with dense and sparse kernels. The data instances had 1,000 dimensions, with five per cent of the elements being nonzero.}
\label{sparsecomparison}
\end{figure}

Compared to the R package, even the CPU version is at least ten times faster. The difference increases with the data size, indicating serious overhead problems in the R implementation.

The GPU results show at least a two-times speedup over the CPU version. This is less than expected, but this result considers a single GPU in the dual configuration, and the CPU is also a higher end model. These results, nevertheless, show that the Thrust template library is not efficient on two-dimensional data structures.

\begin{figure}[t!]
  \centering
  \includegraphics[width=0.47\textwidth, trim = 0 5 0 10, clip]{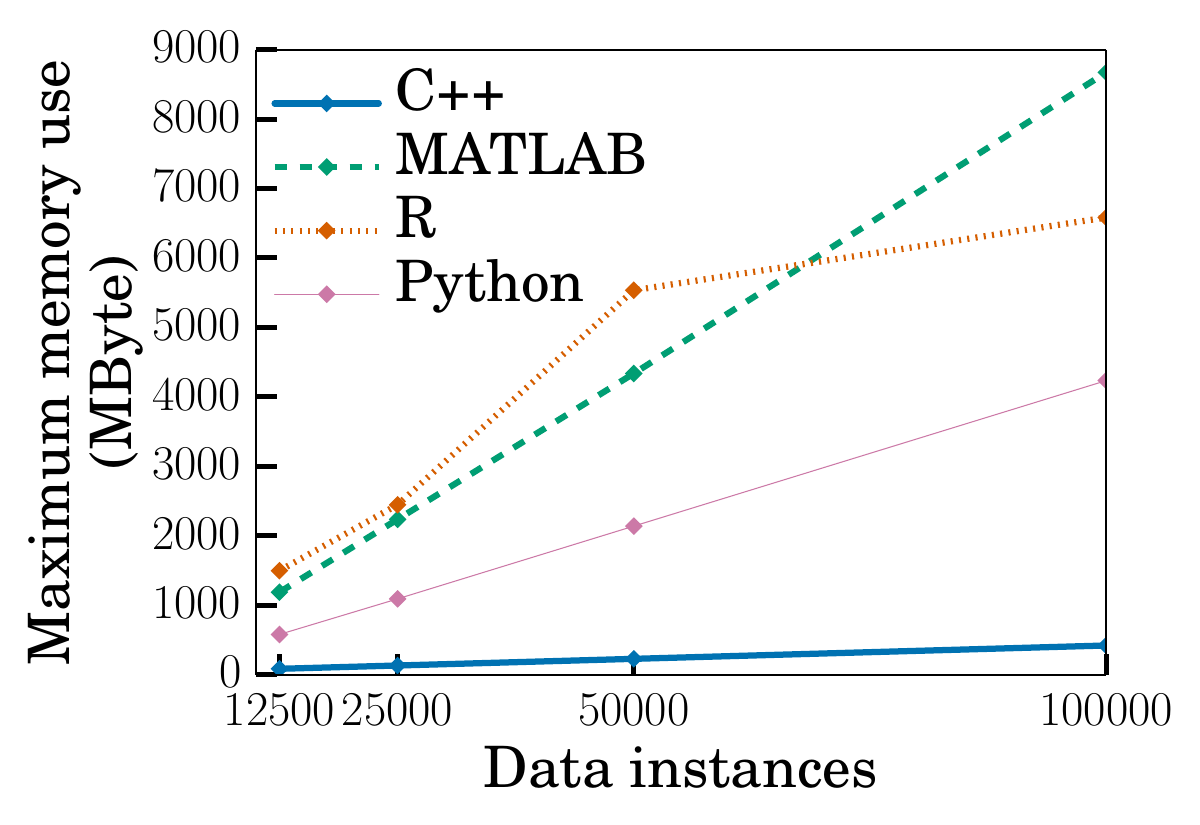}
  \caption{Memory overhead of the Python, R, and MATLAB interfaces compared to the command-line version (indicated as C++).}
  \label{memory_overhead}
\end{figure}

Comparing the sparse and dense kernels on a $50\times 50$ map, we benchmarked with random data instances of 1,000 dimensions that contained five per cent of nonzero elements (Figure~\ref{sparsecomparison}). Execution time was about two times faster with the sparse kernel. The reduction in memory use was far more dramatic, the sparse kernel using only twenty per cent of the memory of the dense one with 100,000 instances. Naturally, the difference with emergent maps would be less apparent, as the code book is always stored in a dense format.

Measuring the memory overhead of the interfaces compared to the native version, we are not surprised to see that the Python variant is the closest (Figure~\ref{memory_overhead}). As the data structures are not duplicated in this interface, it is still counter-intuitive that the gap increases with larger data sets. The R and MATLAB versions have predictably larger and growing gaps compared to the command-line interface, as they both must duplicate all data structures. Apart from the time spent on duplicating the data structures in the R and MATLAB versions, the computational overhead is negligible in all interfaces.

\subsection{Multi-node scaling}
\begin{figure}[t!]
  \centering
  \includegraphics[width=0.47\textwidth, trim = 0 5 0 10, clip]{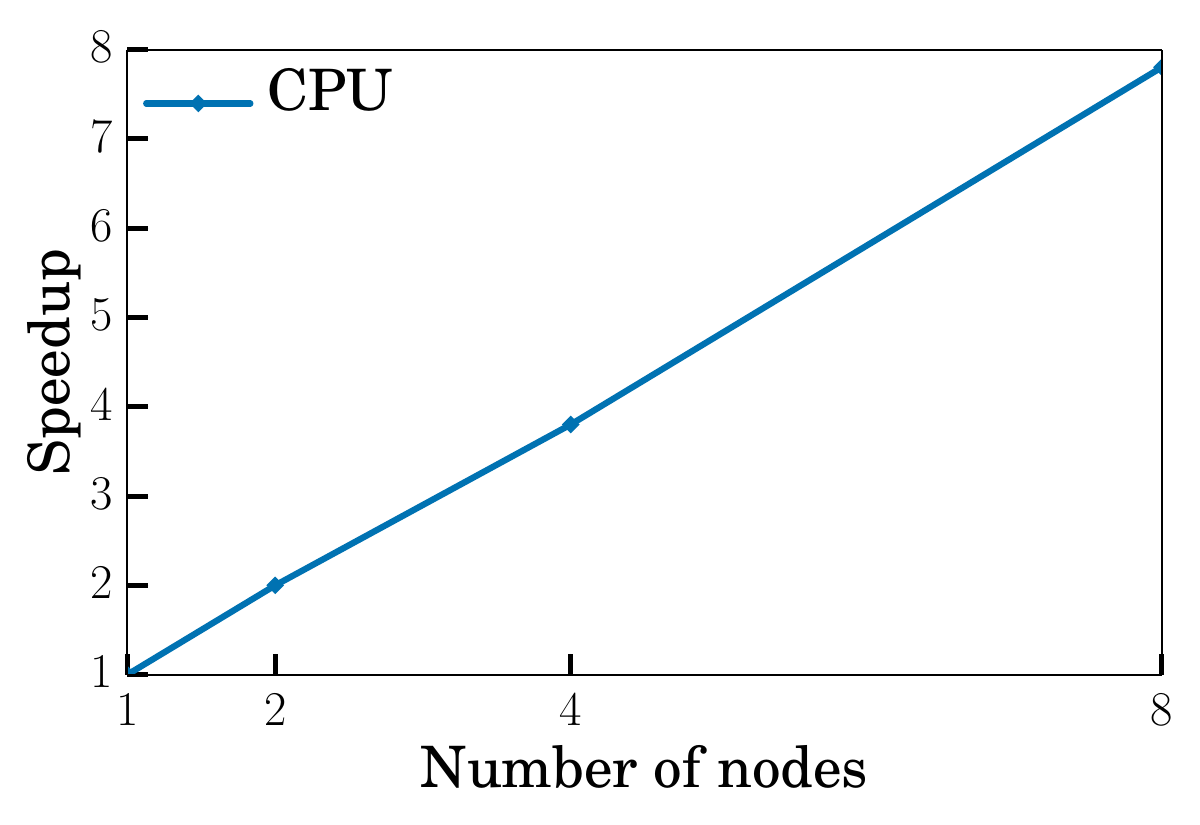}
  \caption{Speedup on multiple nodes with CPU kernel compared to a single node. The data instances had 1,000 dimensions.}
  \label{multinode}
\end{figure}

Using 100,000 instances and a map of $50\times 50$ nodes, the calculations scale in a linear fashion (Figure~\ref{multinode}). This was expected, as there is little communication between nodes, apart from the weight updates. As calculations cannot overlap with communication, we did not benchmark the GPU kernel separately, as its scaling is identical to that of the CPU kernel.

\subsection{Visualization on real data}\label{visualizationresult}
We used the Reuters-21578 document collection for an example on text
mining visualization~\cite{lewis1999rtc}. We used Lucene 3.6.2
\cite{lucene} to create an inverted index of the document
collection. Terms were stemmed and we discarded those that occurred
less than three times or were in the top ten per cent most frequent
ones. Thus we had 12,347 index terms, lying in an approximately
twenty-thousand dimensional space.
\begin{figure}[t!]
  \centering
  \includegraphics[width=\columnwidth, trim = 0 5 0 25, clip]{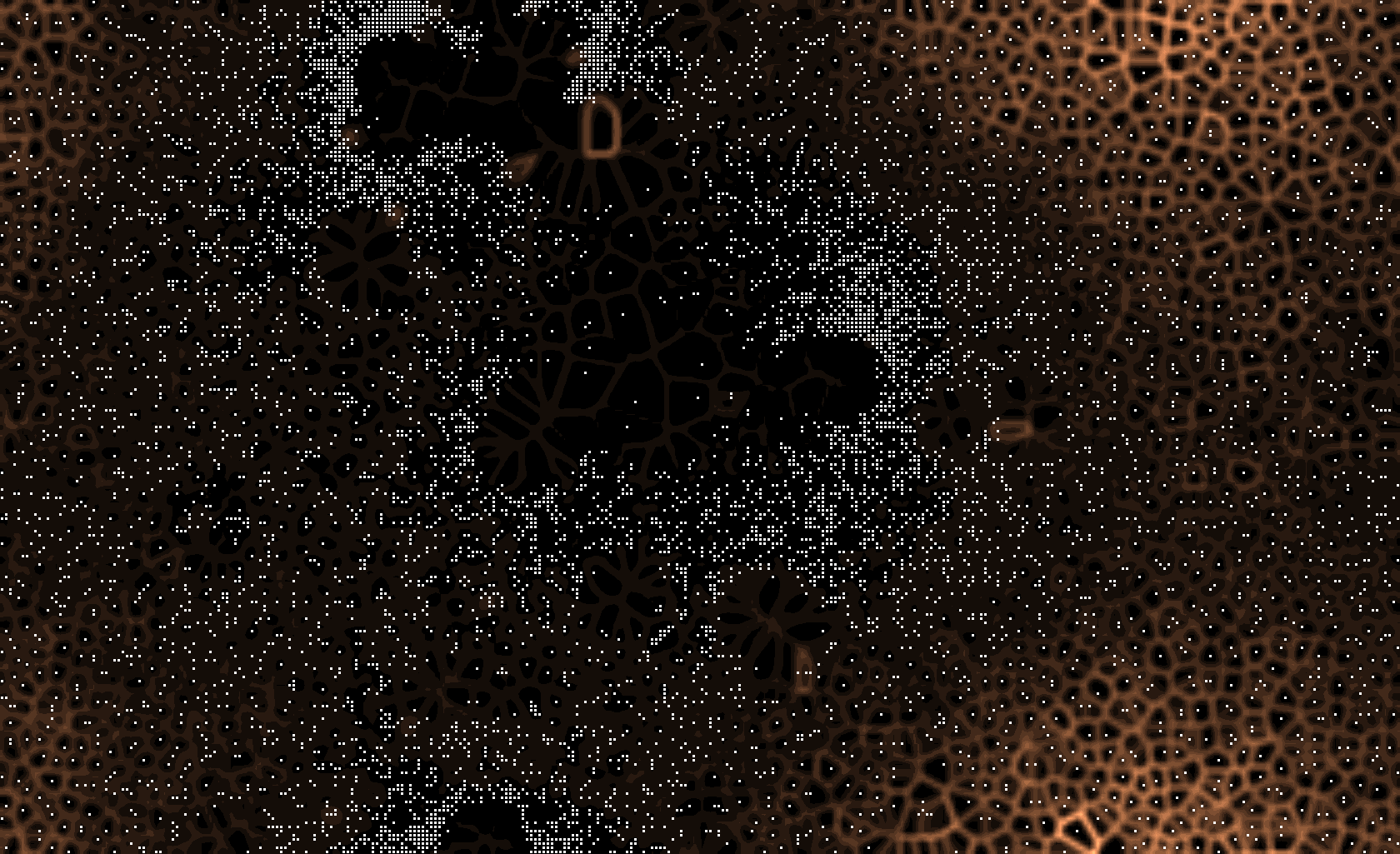}
  \caption{The U-matrix of a toroid emergent self-organizing map after ten epochs of training on the feature space of sparse data. The individual dots are neurons with a weight vector that match a data instance. The other neurons reflect the distances in the original high-dimensional space.}
  \label{sparsesom}
\end{figure}
We trained a toroid emergent self-organizing map of $336\times 205$ dimensions. The initial learning rate was 1.0, which decreased linearly over ten epochs to 0.1. The initial radius for the neighborhood was a hundred neurons, and it also decreased linearly to one. The neighborhood function was a noncompact Gaussian.

We studied the U-matrix of the map. Visualizing this with the
Databionic ESOM Tools~\cite{ultsch2005esom}, we plotted the
global structure of the map in Figure~\ref{sparsesom}. The map clearly
shows dense areas where index terms are close and form tight
clusters. Other parts of the map are sparse, with large barriers
separating index terms into individual semantic regions.

\section{Limitations}
The most constraining limitation is the storing of the code book in
the main memory. While memory use has been optimized, and only the
number of computing nodes sets a limit to the amount of data to be
processed, each node keeps a full copy of the code book. This is not a
problem for feature spaces of a few thousand dimensions -- even
emergent maps are easy to compute. Yet, if the feature space has over
tens of thousands or more features, emergent maps are no longer
feasible.

Currently only the command-line interface is able to unlock all the
capabilities of the computing core. The R,
Python and MATLAB interfaces can use
capabilities of the OpenMP and CUDA core, but these wrappers
cannot use a distributed cluster for the calculations.

The R interface cannot support the GPU kernel easily on
Windows due to incompatibilities between the R supported
compiler GCC and the CUDA supported compiler
Visual C++, while on Linux and OS X it can be built easily.

\section{Conclusions}
Do we need another implementation of self-organizing maps? We believe
the answer is yes. Libraries in high-level languages do not have a
scalable training module, and even implementations on parallel and
distributed architectures could be improved on. Our solution scales
from a single-thread execution to massively parallel GPU-accelerated
clusters using a common core. Batch processing is encouraged by a
command-line interface, whereas interactive use is enabled by
Python, R, and MATLAB
interfaces. Even when called interactively, execution is parallelized.

\begin{itemize}
  \item Memory-efficient multicore execution.
  \item Purely MPI-based, highly efficient distributed implementation.
  \item Optimized hybrid CPU-GPU kernel for boosting training on dense data.
  \item Sparse kernel.
  \item Common computational back-end for popular data-processing languages: Python, R, and MATLAB.
  \item Windows, OS X, Linux support of the command-line interface, with full parallel capability.
  \item Highly efficient on large emergent maps.
  \item Format compatibility with ESOM Tools for visualization.
\end{itemize}

As emergent maps are especially computationally intensive, we believe
that the primary usage scenario is a single-node, multi-GPU configuration
for data exploration, which was only feasible for small maps and small
volumes of data until now. Larger volumes of data are also easy to
compute on a cluster.

\section*{Acknowledgment}
The first author was supported by the European Commission Seventh Framework Programme under Grant Agreement Number FP7-601138 PERICLES and by the AWS in Education Machine Learning Grant award.

\end{document}